\documentclass[letterpaper, 10 pt, conference]{ieeeconf}  

\IEEEoverridecommandlockouts                              
\usepackage{xcolor,amsmath,amssymb,mathtools,booktabs,tikz,pgfplots}

\newtheorem{remark}{Remark}

\pgfplotsset{every axis/.append style={
		label style={font=\large},
		tick label style={font=\large}  
}}
\usepackage{amsmath,mathtools,tikz,booktabs,bm,multirow,pgfplots}
\usepackage[implicit = false]{hyperref}
\usepackage{grffile}

\pgfplotsset{compat=newest}
\usetikzlibrary{plotmarks}
\usetikzlibrary{arrows.meta}
\usepgfplotslibrary{patchplots}

\definecolor{mycolor1}{rgb}{0.00000,0.44700,0.74100}%
\definecolor{mycolor2}{rgb}{0.85000,0.32500,0.09800}%
\definecolor{mycolor3}{rgb}{0.46600,0.67400,0.18800}%

\newcommand{\cov}{\mathrm{cov}}
\newcommand{\Exp}{\mathbb{E}}

\newcommand{\xv}{\mathbf{x}}
\newcommand{\rv}{\mathbf{r}}
\newcommand{\wv}{\mathbf{w}}

\newcommand{\fv}{\mathbf{f}}

\newcommand{\mv}{\mathbf{m}}

\title{\LARGE \bf Gaussian Processes with Noisy Regression Inputs for Dynamical Systems}

\author{Tobias M. Wolff, Victor G. Lopez, and Matthias A. Müller 
\thanks{This project has received funding from the European Research
	Council (ERC) under the European Union’s Horizon 2020 research
	and innovation programme (grant agreement No 948679).}
\thanks{Tobias M. Wolff, Victor G. Lopez, and Matthias A. Müller are with the Leibniz University Hannover, Institute of Automatic Control, 30167 Hannover, Germany
        {\tt\small \{wolff,lopez,mueller\}@irt.uni-hannover.de}}%
}
\begin{document}

\maketitle
\thispagestyle{empty}
\pagestyle{empty}

\begin{abstract}
	This paper is centered around the approximation of dynamical systems by means of Gaussian processes. To this end, trajectories of such systems must be collected to be used as training data. The measurements of these trajectories are typically noisy, which implies that both the regression inputs and outputs are corrupted by noise. However, most of the literature considers only noise in the regression outputs. In this paper, we show how to account for the noise in the regression inputs in an extended Gaussian process framework to approximate scalar and multidimensional systems. We demonstrate the potential of our framework by comparing it to different state-of-the-art methods in several simulation examples. 
\end{abstract}

\section{INTRODUCTION}
\label{sec:introduction}
The application of Gaussian process (GP) regression in the context of dynamical systems has received a substantial interest in recent years. It has been applied for a variety of applications such as, e.g., control \cite{hewing2019cautious,beckers2019stable,Maiworm2021} and state estimation \cite{ko2009gp,buisson2021joint,wolff2024gaussian}. The most common setup for GP regression considers two major assumptions on the measured data. First, it is assumed that the available regression input data are noise-free. Second, the measured regression output data are assumed to be corrupted by independent and identically distributed (iid) Gaussian noise. 

One frequently applied approach to approximate dynamical systems by GPs is to model each component of the transition function $f$ by the posterior means of independently learned GPs \cite{hewing2019cautious,beckers2019stable,ko2009gp,wolff2024gaussian}. To approximate these functions, it is assumed that the states (together with the control inputs) can be measured. Subsequently, the control input and state trajectory are used as regression input data, and the (by one time instant shifted) state trajectory is used as regression output data.

In most practical applications, the state measurements are corrupted by noise. On the one hand, this implies that the regression outputs are corrupted by noise, which is in accordance to the standard GP setting. On the other hand, this entails that the regression inputs are also corrupted by noise, which is \emph{not} covered by the standard GP setting\footnote{Throughout this paper, the notion standard Gaussian processes refers to exact Gaussian process regression for noise-free training inputs.}.

To cope with regression input noise in GP regression, one can use heteroscedastic GPs \cite{kersting2007most,goldberg1997regression} (where a second GP is used to model the noise variance and rather large amounts of data are needed \cite{mchutchon2011gaussian}) or variational methods \cite{titsias2010bayesian,doerr2018probabilistic}. An alternative, which is simple but very effective, has been proposed by \cite{mchutchon2015nonlinear,mchutchon2011gaussian}. The key idea is to propagate the input noise to the output by using first order Taylor approximations of the posterior means (see Section~\ref{sec:preliminaries} below for the details). In \cite{mchutchon2015nonlinear}, the authors show that this approach can outperform variational methods, heteroscedastic GPs and standard GPs.

Our work can be considered as an extension of the framework suggested in \cite{mchutchon2015nonlinear} to dynamical systems. Here, one major difference is that one cannot arbitrarily sample training data points to set up a GP. Instead, one typically can only collect trajectories. We show that these trajectories induce correlations that must be taken into account when setting up a GP to correctly represent dynamical systems. Alongside these theoretical derivations, we illustrate the performance of our proposed extension by means of several simulation examples and compare it to the cases where a dynamical system is directly approximated using the method proposed in \cite{mchutchon2015nonlinear} and standard GP regression \cite{rasmussen2006gaussian}. 

\section{PRELIMINARIES AND PROBLEM SETTING}
\label{sec:preliminaries}
The set of real numbers is denoted by $\mathbb{R}$. The identity matrix of dimension $N$ is denoted by $I_N$. A diagonal matrix with $q_1, \dots, q_n$ on its diagonal entries is denoted by $\mathrm{diag}(q_1, \dots, q_n)$. We denote the Kronecker product by $\otimes$. We denote scalars by small letters, vectors by small bold letters and matrices by capital letters. A vector of zeros of length $n$ is denoted by $\mathbf{0}_n$. A square matrix of zeros of dimension $n$ is denoted by $0_{n\times n}$.
\subsection{Exact GP regression}
We briefly review the fundamentals of standard Gaussian processes; a more detailed introduction to GPs can be found in \cite{rasmussen2006gaussian}. GPs are commonly applied to approximate some nonlinear function $\bar{f}: \mathbb{R}^{\bar{n}} \rightarrow \mathbb{R}$. They are fully described by a mean function $m: \mathbb{R}^{\bar{n}} \rightarrow \mathbb{R}$ and a covariance function (also referred to as kernel) $k: \mathbb{R}^{\bar{n}} \times \mathbb{R}^{\bar{n}}\rightarrow \mathbb{R}$. For some $\bar{\xv}, \bar{\xv}' \in \mathbb{R}^{\bar{n}}$, we write
\begin{align}
	\bar{f}(\bar{\xv}) \sim \mathcal{GP}(m(\bar{\xv}),k(\bar{\xv},\bar{\xv}'))
\end{align}
to denote that the function $\bar{f}$ follows a GP with mean function $m$ and covariance function $k$. We collect $N$ regression input and output data points from the unknown function and use them to define $\bar{X} = \begin{pmatrix}
	\bar{\xv}(0) & \dots & \bar{\xv}(N-1)
\end{pmatrix}$ and $\bar{\textbf{Y}} = \begin{pmatrix}
	\bar{y}(0) & \dots & \bar{y}(N-1)
\end{pmatrix}^\top$, respectively. The regression outputs are given by $\bar{y} = \bar{f}(\bar{\xv}) + \bar{\varepsilon}$ with $\bar{\varepsilon}$ being iid Gaussian noise with zero mean and variance $\sigma_{\bar{\varepsilon}}^2$. The key idea of Gaussian process regression is to condition the prior distribution on the training data, which results in a posterior distribution. For some test input $\bar{\mathbf{x}}_{\ast}$, the mean and variance of the posterior distribution are given by \cite[Ch. 2]{rasmussen2006gaussian}
\begin{align}
	&\bar{m}_+(\bar{\xv}_{\ast}|\bar{X}, \bar{\textbf{Y}}) = \mathbf{k}(\bar{\xv}_{\ast},\bar{X})(K(\bar{X},\bar{X}) + \sigma_{\bar{\varepsilon}}^2I_N)^{-1}\bar{\mathbf{Y}} \label{def:pos:mean} \\
	&\bar{\sigma}_+^2(\bar{\xv}_{\ast}|\bar{X},\bar{\textbf{Y}}) = \nonumber\\  &\mathbf{k}(\bar{\xv}_{\ast},\bar{\xv}_{\ast})- \mathbf{k}(\bar{\xv}_{\ast},\bar{X}) (K(\bar{X},\bar{X}) + \sigma_{\bar{\varepsilon}}^2I_N)^{-1}\mathbf{k}(\bar{X},\bar{\xv}_{\ast}), \label{def:post:var}
\end{align}
for $\mathbf{k}(\bar{\xv}_{\ast},\bar{X}) = \begin{pmatrix}
	k(\bar{\xv}_{\ast},\bar{\xv}_i)
\end{pmatrix}_{\bar{\xv}_i \in \bar{X}} = \mathbf{k}(\bar{X},\bar{\xv}_{\ast})^\top$, with~$\mathbf{k}(\bar{\xv}_{\ast},\bar{X}) \in \mathbb{R}^{1 \times N}$, and $K(\bar{X},\bar{X}) = (k(\bar{\xv}_i,\bar{\xv}_j))_{\bar{\xv}_i, \bar{\xv}_j \in \bar{X}}$ with~$K(\bar{X},\bar{X}) \in \mathbb{R}^{N \times N}$. The kernel depends on hyperparameters (such as, e.g., the signal variance and the length scales in case of the squared exponential kernel) that are commonly determined by maximizing the log marginal likelihood, see, e.g., \cite[Eq. (2.30)]{rasmussen2006gaussian}.

\subsection{GP regression with noisy training inputs}
\label{sec:standard:GP:noisy}
These standard results in Gaussian processes rely on the assumption that the regression input data are noise-free. In turn, if the regression input data points are affected by some noise such that only 
\begin{equation}
	\label{def:noisy_regression_input}
	\check{\bar{\xv}} \coloneqq \bar{\xv} + \bar{\rv}	
\end{equation}
is available with $\bar{\rv}$ being some iid Gaussian noise with variance $\Sigma_{\bar{r}} = \mathrm{diag}(\sigma^2_{\bar{r}}, \dots, \sigma^2_{\bar{r}})$, we cannot use standard GP tools anymore, since the problem of exact GP regression based on noisy regression inputs is intractable \cite[Sec. 2.3.2]{deisenroth2010efficient}. We here briefly review the work of \cite[Ch. 2]{mchutchon2015nonlinear} (which is more detailed than the original work \cite{mchutchon2011gaussian}) to handle this issue. First, a Taylor series expansion around the noisy regression input is done (and truncated after the first-order term), which results in
\begin{align}
	\bar{f}(\bar{\xv})= \bar{f}(\check{\bar{\xv}}-\bar{\rv}) \approx \bar{f}(\check{\bar{\xv}}) - \frac{\partial \bar{f}(\xv)}{\partial \xv} \Big\rvert_{\xv = \check{\bar{\xv}}}\bar{\rv}.
\end{align}
The second term depends on the derivative of a GP, which is again a GP \cite{Solak2002}. Although one can compute the first and second moment of this expression, it is much simpler to perform another approximation by replacing the derivative of the GP by the derivative of its posterior mean \cite{mchutchon2015nonlinear}. In this case, we consider the following model
\begin{align}
	\label{model:mchutchon}
	\bar{y} \approx \bar{f}(\check{\bar{\xv}}) - \frac{\partial \bar{m}_+(\bar{\xv}|\check{\bar{X}},\bar{Y})}{\partial \bar{\xv}} \Big\rvert_{\bar{\xv} = \check{\bar{\xv}}}\bar{\rv} + \bar{\varepsilon}.
\end{align}
This model results in the following approximate covariance matrix of the training targets
\begin{align}
	\check{K} = &\begin{pmatrix}
		k(\check{\bar{\xv}}(0), \check{\bar{\xv}}(0))  & \dots  & k(\check{\bar{\xv}}(0), \check{\bar{\xv}}(N-1))\\
		\vdots & \ddots & \vdots \\
		k(\check{\bar{\xv}}(N-1), \check{\bar{\xv}}(0))& \dots & k(\check{\bar{\xv}}(N-1), \check{\bar{\xv}}(N-1))
	\end{pmatrix} \nonumber \\
	&+ \mathrm{diag}(\bar{\sigma}^2_{\mathrm{out}}(0), \dots, \bar{\sigma}^2_{\mathrm{out}}(N-1))
	\label{def:covariance:noisy:inputs}
\end{align}
with
\begin{align}
	&\bar{\sigma}^2_{\mathrm{out}}(i) \coloneqq \nonumber\\ 
	&\frac{\partial \bar{m}_+(\bar{\xv}|\bar{X},\bar{Y})}{\partial \bar{\xv}} \Big\rvert_{\bar{\xv} = \check{\bar{\xv}}(i)} \Sigma_{\bar{r}}\frac{\partial \bar{m}_+(\bar{\xv}|\bar{X},\bar{Y})}{\partial \bar{\xv}} \Big\rvert_{\bar{\xv} = \check{\bar{\xv}}(i)}^\top + \sigma_{\bar{\varepsilon}}^2. 
\end{align}
The expressions of the posterior mean and variance are analogous to~(\ref{def:pos:mean}) and~(\ref{def:post:var}), simply with $K(\bar{X},\bar{X}) + \sigma_{\bar{\varepsilon}}^2I_N$ replaced by $\check{K}$ from (\ref{def:covariance:noisy:inputs}). Note that we have one further hyperparameter to determine, which is the variance of the input noise. The optimization of the hyperparameters must be adapted, since the covariance matrix now depends on the derivatives of the posterior mean. Hence, \cite{mchutchon2015nonlinear} proposes to iterate the computations of the slopes of the posterior mean and the optimization of the hyperparameters. Note that the approach does not differ from a standard GP for (i) negligible input noise levels and (ii) constant posterior mean gradients \cite{mchutchon2015nonlinear}. Finally, in simulation examples this approach often outperforms heteroscedastic GPs, standard GPs, as well as variational methods \cite{mchutchon2015nonlinear}.

\subsection{Problem formulation}
In this work, we focus on discrete-time nonlinear dynamical systems of the following form\footnote{To simplify the notation, we do not consider control inputs in (\ref{def:dynamical:system}). However, the results of this paper can be straightforwardly extended to systems with control inputs.}
\begin{align}
	\label{def:dynamical:system}
	\xv(t+1) &= \fv(\xv(t)) + \wv(t)
\end{align}
with states $\xv\in \mathbb{R}^n$, process noise $\wv\in \mathbb{R}^n$ (sometimes also referred to as system noise), and $\fv:\mathbb{R}^n \rightarrow \mathbb{R}^n$. The process noise $\wv$ is assumed to be iid Gaussian noise with zero mean and variance $\Sigma_w = \mathrm{diag}(\sigma_w^2, \dots, \sigma_w^2)$. Here, we assume the same noise variance among all components to simplify the analysis. The objective of this work is to approximate the function $\fv$ by (the posterior means of) Gaussian processes. To this end, we collect a sufficiently long (or multiple shorter) trajectory from the system. In the here considered setting of dynamical systems, we cannot collect arbitrary data points. This is due to the recursive structure of~(\ref{def:dynamical:system}): the (noisy) outputs of the function $\fv$ at some time instant correspond to the function inputs at the next time instant.

\begin{figure*}[!t]
	\begin{align}
		\tilde{K} = \begin{pmatrix}
			k(\tilde{x}(0), \tilde{x}(0)) + \sigma^2_{\mathrm{out}}(0) & k(\tilde{x}(0), \tilde{x}(1)) - \nabla_0 \sigma^2_{r}& \dots  & k(\tilde{x}(0), \tilde{x}(N-1)) \\
			k(\tilde{x}(1), \tilde{x}(0)) - \nabla_0 \sigma^2_{r} & k(\tilde{x}(1), \tilde{x}(1)) + \sigma^2_{\mathrm{out}}(1) & \dots  & k(\tilde{x}(1), \tilde{x}(N-1)) \\	
			\vdots & \vdots & \ddots & \vdots \\
			k(\tilde{x}(N-2), \tilde{x}(0)) & k(\tilde{x}(N-2), \tilde{x}(1)) &  &  k(\tilde{x}(N-2), \tilde{x}(N-1)) - \nabla_{N-2} \sigma^2_{r} \\
			k(\tilde{x}(N-1), \tilde{x}(0))  & k(\tilde{x}(N-1), \tilde{x}(1)) & \dots & k(\tilde{x}(N-1), \tilde{x}(N-1)) + \sigma^2_{\mathrm{out}}(N-1)  
		\end{pmatrix}
		\label{def:covariance:off:diagonal}
		\tag{$\star$}
	\end{align}
	\hrule
\end{figure*}

When measuring a trajectory from the system, one has (in most applications) only access to noisy measurements of the trajectories (due to, e.g., noise coming from the sensors). This means that only 
\begin{align}
	\tilde{\xv}(0) &= \xv(0) + \rv(0) \label{eq:noisy:initial}\\
	\tilde{\xv}(1) &= \xv(1) + \rv(1) = \fv(\xv(0)) + \wv(0) + \rv(1)\\
	&\: \: \:\vdots \nonumber \\
	\tilde{\xv}(N) &=  \fv(\xv(N-1)) + \wv(N-1) + \rv(N) \label{eq:noisy:end}
\end{align}
can be measured with $\rv$ being iid Gaussian noise with variance $\Sigma_r = \mathrm{diag}(\sigma_r^2, \dots, \sigma_r^2)$. Note that we consider some measurement noise $\rv$ \emph{in addition} to the standard process noise $\wv$ (which is often considered in the context of GP based control and estimation, compare, e.g., \cite{hewing2019cautious,ko2009gp}). The measurement noise $\rv$ and the process noise $\wv$ are assumed to be independent. To approximate the function~$\fv$, we have $\tilde{\xv}(0), \dots, \tilde{\xv}(N-1)$ as regression input data and $\tilde{\xv}(1), \dots, \tilde{\xv}(N)$ as regression output data available. We do not have access to the true regression inputs, i.e., $\xv(0), \dots, \xv(N-1)$. The subject of this work is to propose a framework to account for the input noise in the case of dynamical systems, where only noisy trajectories are available as training data.

\section{SCALAR SYSTEMS}
\label{sec:input:noise:scalar}
\subsection{Analysis of regression input noise}
In this section, we consider $f: \mathbb{R} \rightarrow \mathbb{R}$ and $x\in \mathbb{R}$. As training data, we assume that one trajectory of length $N+1$ has been collected to set up the GP. We use the same approach as in (\ref{model:mchutchon}) and introduce 
\begin{align}
	\nabla_i \coloneqq \frac{\partial m_+(x|\tilde{\mathbf{X}}^{\mathrm{in}}, \tilde{\mathbf{X}}^{\mathrm{out}})}{\partial x} \Big\rvert_{x = \tilde{x}(i)}
\end{align}
with $\tilde{\mathbf{X}}^{\mathrm{in}} = \begin{pmatrix}
	\tilde{x}(0) & \dots & \tilde{x}(N-1)
\end{pmatrix}$ and $\tilde{\mathbf{X}}^{\mathrm{out}} = \begin{pmatrix}
	\tilde{x}(1) & \dots & \tilde{x}(N)
\end{pmatrix}$ to denote the derivative of the posterior mean approximating the function $f$ at the location $\tilde{x}(i)$ (which is here denoted by $m_+$ instead of $\bar{m}_+$ since we consider dynamical systems and not standard functions as in Section~\ref{sec:standard:GP:noisy}). Together with (\ref{eq:noisy:initial}) - (\ref{eq:noisy:end}), this results in 
\begin{align*}
	\tilde{x}(i) \approx f(\tilde{x}(i-1)) - \nabla_{i-1}r(i-1) + w(i-1) + r(i).
\end{align*}
The variance corresponds to 
\begin{align}
	&\cov\big(\tilde{x}(i), \tilde{x}(i)\big) \nonumber  \\
	&\approx  \mathbb{E}\bigg\{\Big(f(\tilde{x}(i-1)) - \nabla_{i-1} r(i-1) + w(i-1) + r(i) \nonumber \\
	& - \mathbb{E}\big\{f(\tilde{x}(i-1)) - \nabla_{i-1} r(i-1) + w(i-1) + r(i)\big\}\Big)^2\bigg\}  \nonumber \\
	&=  k(\tilde{x}(i-1), \tilde{x}(i-1)) + \nabla_{i-1} \Sigma_{r}\nabla_{i-1} + \sigma_{w}^2 + \sigma_r^2 \nonumber \\
	&\eqqcolon k(\tilde{x}(i-1), \tilde{x}(i-1)) + \sigma^2_{\mathrm{out}}(i-1)
\end{align}
for all $i = 1, \dots, N$, since (i) $\wv$ and $\rv$ are independent and (ii) $\rv$ is assumed to be iid. 

We compute the covariance of two subsequent samples
\begin{align*}
	&\cov\big(\tilde{x}(i+1), \tilde{x}(i)\big)\approx \nonumber\\
	&\mathbb{E}\bigg\{\Big(f(\tilde{x}(i)) - \nabla_{i} r(i) + w(i) + r(i+1) \nonumber \\ 
	& - \mathbb{E}\big\{f(\tilde{x}(i)) - \nabla_i r(i) + w(i) + r(i+1)\big\}\Big) \nonumber \\
	&\hspace{0.5cm} \Big(f(\tilde{x}(i-1)) - \nabla_{i-1} r(i-1) + w(i-1) + r(i) \nonumber \\
	& - \mathbb{E}\big\{f(\tilde{x}(i-1)) - \nabla_{i-1} r(i-1)+ w(i-1) + r(i)\big\}\Big)\bigg\}, 
\end{align*}
 resulting in
\begin{align}
	\cov\big(\tilde{x}(i+1), \tilde{x}(i)\big)&\approx k(\tilde{x}(i), \tilde{x}(i-1)) - \Exp\big\{\nabla_i r(i)r(i)\big\} \nonumber \\
	&= k(\tilde{x}(i), \tilde{x}(i-1)) - \nabla_i \sigma_{r}^2 \label{eq:cov:scalar}
\end{align}
for all $i = 1, \dots, N-1$ and similarly for $\cov\big(\tilde{x}(i), \tilde{x}(i+1)\big)$. 
The term $- \nabla_i \sigma_{r}^2$ in (\ref{eq:cov:scalar}) appears only in the covariance of two consecutive data points (i.e., $x(i)$ and $x(i+1)$) and is caused by the recursive nature of (\ref{def:dynamical:system}) and the propagation of the input noise to the output in (\ref{model:mchutchon}). For this reason, this term does not appear in the developments in \cite{mchutchon2015nonlinear}, where dynamical systems are not the central focus. In our case, the covariance matrix of the measured data corresponds to the expression given in (\ref{def:covariance:off:diagonal}) above, where the term $- \nabla_i \sigma_{r}^2$ appears only in the entries immediately above and below the main diagonal.

If one does not consider consecutive samples in the training data, the additional term in (\ref{eq:cov:scalar}) vanishes. In the context of dynamical systems, this could happen if (i) one only uses every second data point (which could decrease the performance since half of the data are lost) or (ii) one performs one-step experiments, such that a regression output does not become a regression input. Intuitively, this means that one considers some initial condition, measures the next state and then considers a different initial condition, which is not meaningful/possible for many applications. Note that our above theoretical analysis focuses on collecting one single trajectory. If one considers multiple trajectories, the entries in the covariance matrix describing the transition from one trajectory to another do not contain the additional term~$- \nabla_i \sigma^2_{r}$.

The last step is to set up the posterior mean and the posterior variance, which is once again analogous to~(\ref{def:pos:mean}) and~(\ref{def:post:var}) with $K(\bar{X},\bar{X}) + \sigma_{\bar{\varepsilon}}^2$ replaced by $\tilde{K}$ from (\ref{def:covariance:off:diagonal}). 

\subsection{Application to logistic growth example}

\begin{figure}[t!]
	\centering
	\vspace{0.125cm}
	\includegraphics[scale=0.35]{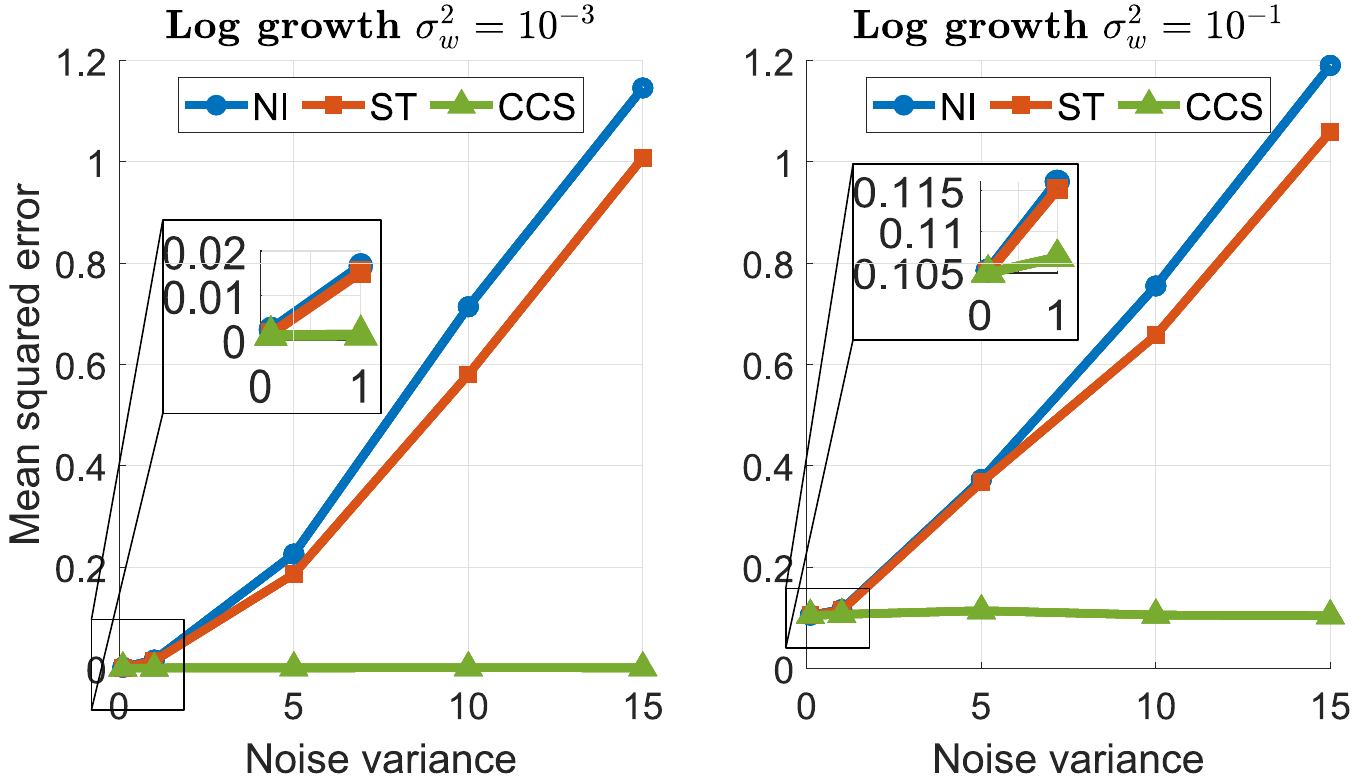}
	\caption{Simulation results of example (\ref{def:log:growth}) considering two different process noise variances as indicated in the titles of the plots. We implement the here proposed extension (referred to as ``CCS" standing for ``covariance of consecutive samples"), a standard GP (called ``ST") and the approach proposed by \cite{mchutchon2015nonlinear} (called ``NI" standing for ``noisy inputs", which is the abbreviation given by the authors in \cite{mchutchon2015nonlinear} to describe their framework). We report the MSE as defined in (\ref{def:mse}), respectively.}
	\label{fig:1d}
\end{figure}

We evaluate the effect of the additional off-diagonal terms for a numerical example\footnote{The code of the simulations is available here: \url{https://doi.org/10.25835/xwkni4f6}}. We use a zero prior mean and a squared exponential kernel. We consider the following (Euler-discretized) system 
\begin{align}
	\label{def:log:growth}
	x(k+1) = x(k) + Tqx(k)\big(1-x(k)/C\big) + w(k)
\end{align}
with $T= 1, q= 0.1, C = 100$, which corresponds to a logistic growth example \cite{Tsoularis2002}. Note that the relatively small value of $T$ and the rather large value of $C$ imply that we only have to deal with a small nonlinearity and almost constant gradients. We collect three trajectories of length 100. We consider normally distributed process noise with mean $\mu_w = 0$ and variance $\sigma_w^2 = 10^{-3}$ (and in a second run normally distributed process noise with $\mu_w = 0$ and variance $\sigma_w^2 = 10^{-1}$) as well as normally distributed measurement noise with mean $\mu_r = 0$ and various variances as illustrated in Figure~\ref{fig:1d}. We use five iterations of slope/hyperparameter computations, compare \cite{mchutchon2015nonlinear}. To test the performance of the GPs, we consider $N_{\ast} = 500$ random samples from a uniform distribution $\mathcal{U}(0,100)$ and compute the posterior mean. We compare our method to the one proposed by \cite{mchutchon2015nonlinear} and to a standard GP \cite{rasmussen2006gaussian}. In all cases, we then compute the mean squared error (MSE) defined as
\begin{align}
	\text{MSE} \coloneqq \frac{1}{N_{\ast}} \sum_{k=1}^{N_{\ast}} ||f(x_{\ast}(i)) - m_+(x_{\ast}(i)|\tilde{\mathbf{X}}^{\mathrm{in}}, \tilde{\mathbf{X}}^{\mathrm{out}})||^2 \label{def:mse}.
\end{align}
The simulation results are displayed in Figure~\ref{fig:1d}. We observe that our proposed extension substantially outperforms the other approaches for both process noise variances, in particular for large measurement noise variances. This is due to the explicit consideration of the covariance between two consecutive samples, which is not considered in the framework from \cite{mchutchon2015nonlinear} and a standard GP \cite{rasmussen2006gaussian}. 

When a larger process noise variance is considered (compare Fig. \ref{fig:1d}, right plot), our proposed approach still outperforms the other two, although the difference becomes slightly smaller. This is due to the fact that in this case, the diagonal terms in the covariance matrix become more dominant and the advantage of our approach (that considers the noise variance in the entries immediately above and below the main diagonal) becomes less prominent.

Moreover, we observe that the performance of the standard GP and the method proposed by \cite{mchutchon2015nonlinear} is similar (with a slight advantage for the standard GP). This is due to the considered system which is almost linear. In this case, the gradients of the posterior mean are almost constant and a standard GP can achieve a similar effect (than the extension of \cite{mchutchon2015nonlinear}) by simply increasing the noise variance. The slight advantage for the standard GP may be due to a minor overfitting in case of the approach proposed by \cite{mchutchon2015nonlinear}, where we have one more hyperparameter to determine.

\section{MULTIDIMENSIONAL SYSTEMS}
\label{sec:input:noise:multi}
\subsection{Analysis of regression input noise}
In this section, we now focus on multidimensional systems. This means that we consider some function $\mathbf{f}: \mathbb{R}^n \rightarrow \mathbb{R}^n$ with $\xv \in \mathbb{R}^n$. The most common approach to approximate these systems is to consider the individual components of the function~$\fv$ to be independent \cite{hewing2019cautious,beckers2019stable,ko2009gp}. In this case, scalar GPs are used to approximate each component of the function~$\fv$. Alternatively, one can use a linear model of coregionalization \cite{alvarez2012kernels}, where all components are learned jointly (and hence also correlations among the components can be learned).

\begin{figure*}[t!]
		\begin{minipage}[c]{0.7\textwidth}
			\centering
			\vspace{0.125cm}
			\includegraphics[scale=0.3]{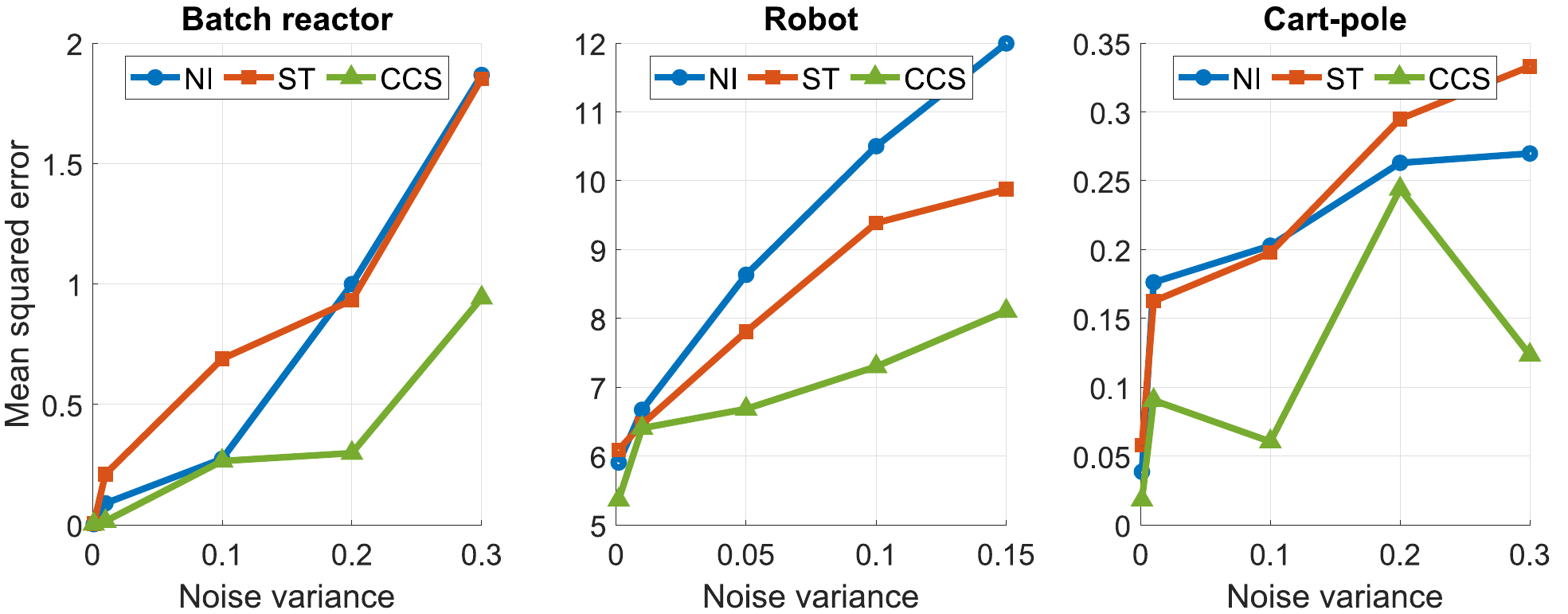}
		\end{minipage}
		\begin{minipage}[c]{0.3\textwidth}
			\caption{Simulation results for a batch reactor, a two-link planar robot, and a cart pole system. The figures show the performances of the here proposed framework (referred to as ``CCS" standing for ``covariance of consecutive samples"), the extension by \cite{mchutchon2015nonlinear} (called ``NI" standing for ``noisy inputs", which is the abbreviation given by the authors in \cite{mchutchon2015nonlinear} to describe their framework), and a standard GP (called ``ST") for randomly sampled test data. We report the MSE as defined in~(\ref{def:mse}).}
			\label{fig:md}
		\end{minipage}				
\end{figure*}

However, the above works rely on the assumption that the regression input data are noise-free. As mentioned in the previous section, this is rarely the case in the context of dynamical systems, since the measurements of the states are corrupted by some noise. In the following, we again consider the input noise by applying the approach proposed in \cite{mchutchon2015nonlinear} (compare (\ref{model:mchutchon})) to dynamical systems. As shown in the following derivation, analogous to Section~\ref{sec:input:noise:scalar}, we obtain additional terms in the covariance between two consecutive observations. Moreover, in addition to the scalar case, we also obtain covariance terms between the regression outputs corresponding to the different components of $f$. In particular, since
\begin{align}
	\tilde{x}_j(i) \approx& f_j(\tilde{\xv}(i-1)) - \frac{\partial m_{+,j}(\xv|\tilde{X}^{\mathrm{in}},\tilde{X}^{\mathrm{out}})}{\partial \xv}\Big |_{\xv = \tilde{\xv}(i-1)} \times \nonumber\\
	&\textbf{r}(i-1)   + w_j(i-1)  + r_j(i)
\end{align}
for all $j = 1, \dots, n$, we obtain
\begin{align}
	&\mathrm{cov}\big(\tilde{x}_j(i), \tilde{x}_{\ell}(i)\big) \nonumber \\ 
	&\approx \mathbb{E}\Bigg\{\bigg( f_j(\tilde{\xv}(i-1)) - \mathbb{E}\Big\{f_j(\tilde{\xv}(i-1)) \Big\}\bigg) \times\nonumber \\
	&\qquad \bigg( f_{\ell}(\tilde{\xv}(i-1)) - \mathbb{E}\Big\{f_{\ell}(\tilde{\xv}(i-1)) \Big\}\bigg)^\top \Bigg\}\nonumber  \\ 
	&+ \mathbb{E}\Bigg\{ \frac{\partial m_{+,j}(\xv|\tilde{X}^{\mathrm{in}},\tilde{X}^{\mathrm{out}})}{\partial \xv}\Big |_{\xv = \tilde{\xv}(i-1)} \textbf{r}(i-1) \times \nonumber \\
	& \qquad \mathbf{r}(i-1)^\top \frac{\partial m_{+,{\ell}}(\xv|\tilde{X}^{\mathrm{in}},\tilde{X}^{\mathrm{out}})}{\partial \xv}\Big |_{\xv= \tilde{\xv}(i-1)}^\top \Bigg\} \nonumber \\ 
	&+  (\sigma^2_r + \sigma^2_w)\delta_{j\ell}
\end{align}
with $\delta_{\ell,j}$ denoting the Kronecker delta. To simplify the analysis, we assume that the different GPs modeling the different components are mutually independent (as commonly done in the context of GP based control/estimation \cite{hewing2019cautious,beckers2019stable,ko2009gp}). Consequently, it holds that
\begin{align}
	\mathbb{E}\{f_j(\tilde{\mathbf{x}}) f_{\ell}(\tilde{\mathbf{x}})\} = \mathbb{E}\{f_j(\tilde{\mathbf{x}})\} \mathbb{E}\{f_{\ell}(\tilde{\mathbf{x}})\} 
\end{align}
and therefore
\begin{align}
	&\mathrm{cov}\big(\tilde{x}_j(i), \tilde{x}_{\ell}(i)\big) \approx \big(k(\tilde{\mathbf{x}}(i-1),\tilde{\mathbf{x}}(i-1))+\sigma^2_r + \sigma^2_w\big)\delta_{j\ell} 
	 \nonumber \\
	 &+\frac{\partial m_{+,j}(\xv|\tilde{X}^{\mathrm{in}},\tilde{X}^{\mathrm{out}})}{\partial \xv}\Big |_{\xv = \tilde{\xv}(i-1)}   \Sigma_r\times \nonumber \\
	& \hspace{2cm} \frac{\partial m_{+,{\ell}}(\xv|\tilde{X}^{\mathrm{in}},\tilde{X}^{\mathrm{out}})}{\partial \xv}\Big |_{\xv= \tilde{\xv}(i-1)}^\top  \label{cov:components}
\end{align}
for all $i = 1, \dots, N$ and $j,\ell  = 1, \dots, n$. Hence, although assuming independence among the different GPs (modeling the different components), the observations covary due to the input noise. Moreover, as in the previous section (compare (\ref{eq:cov:scalar})), we need to consider the covariance within the same component, but for subsequent time instants

\begin{figure*}[t!]
	\begin{align}
		K_{\mathrm{md}} =& (K_x + \sigma_r^2I_N + \sigma_w^2I_N)\otimes I_{n} + \begin{pmatrix}
			\nabla_0\Sigma_r\nabla_0^\top   & -\nabla_1^\top\sigma_r^2& 0_{n\times n}  & \dots & 0_{n\times n}  \\			
			-\nabla_1\sigma_r^2& \nabla_1\Sigma_r\nabla_1^\top   &-\nabla_2^\top\sigma_r^2& \dots & 0_{n\times n}  \\			
			0_{n\times n}  & -\nabla_2 \sigma_r^2&\nabla_2\Sigma_r\nabla_2^\top   & \dots & 0_{n\times n}  \\				
			\vdots & \vdots & \vdots & \ddots & \vdots \\		
			0_{n\times n} & 0_{n\times n} & 0_{n\times n} & \dots &\nabla_{N-1}\Sigma_r\nabla_{N-1}^\top  \\
		\end{pmatrix}
		\label{def:covariance:matrix:2D}
		\tag{$\star\star$}
	\end{align}
	\hrule
\end{figure*}

\begin{align}
	\mathrm{cov}\big(\tilde{x}_j(i), &\tilde{x}_j(i+1)\big) \approx k(\tilde{\xv}(i-1), \tilde{\xv}(i))  \nonumber\\ 
	&-  \frac{\partial m_{+,j}(\xv|\tilde{X}^{\mathrm{in}},\tilde{X}^{\mathrm{out}})}{\partial x_j}\Big |_{\xv= \tilde{\xv}(i)}\sigma_r^2
\end{align}
for all $j =1, \dots, n$ and $i = 1, \dots, N-1$ and similarly for $	\mathrm{cov}\big(\tilde{x}_j(i+1), \tilde{x}_j(i)\big)$. Finally, we need to consider the covariance between observations of two different (not necessarily adjacent) components and subsequent time instants as, e.g., 
\begin{align*}
	&\mathrm{cov}\big(\tilde{x}_j(i), \tilde{x}_{\ell}(i+1)\big)
	\approx  - \frac{\partial m_{+,{\ell}}(\xv|\tilde{X}^{\mathrm{in}},\tilde{X}^{\mathrm{out}})}{\partial x_j}\Big |_{\xv= \tilde{\xv}(i)} \sigma^2_r.
\end{align*}
for all $j, \ell = 1, \dots, n$ (but $j \neq \ell$) and $i = 1,\dots, N-1$ and similarly for $\mathrm{cov}(\tilde{x}_j(i+1), \tilde{x}_\ell(i))$. 

To set up a GP for this case, we cannot proceed in the standard way by simply learning individual GPs. This is due to the correlations among the different components, compare~(\ref{cov:components}), which cannot be considered by learning the components individually. Instead, we here learn all the components of the GP jointly (which requires multi-output regression implying a higher computational complexity). To this end, we set up the vector of observations
\begin{align}
	&X^\mathrm{out} = \nonumber \\
	&\begin{pmatrix}
		\tilde{x}_1(1) &  \dots& \tilde{x}_n(1)&\tilde{x}_1(2)&\dots & \tilde{x}_n(N)
	\end{pmatrix}^\top.
\end{align}
The covariance matrix corresponds to the expression given in (\ref{def:covariance:matrix:2D}) below with $K_x = (k(\tilde{\mathbf{x}}(i), \tilde{\mathbf{x}}(j)))_{\tilde{\mathbf{x}}(i), \tilde{\mathbf{x}}(j) \in \tilde{X}^{\mathrm{in}}}$ and $\mathbf{\nabla}_i$ 
defined as 
\begin{align*}
\mathbf{\nabla}_i \coloneqq \begin{pmatrix}
		\frac{\partial m_{+,1}}{\partial x_1} |_{\xv = \tilde{\xv}(i)} & 		\frac{\partial m_{+,1}}{\partial x_2} |_{\xv = \tilde{\xv}(i)} & 
		\dots \\
		\frac{\partial m_{+,2}}{\partial x_1} |_{\xv = \tilde{\xv}(i)} & 		\frac{\partial m_{+,2}}{\partial x_2} |_{\xv = \tilde{\xv}(i)} & 
		\dots \\
		\vdots & \vdots & \ddots \\
	\end{pmatrix}.
\end{align*}
The predictive mean and variance are given by 
\begin{align}
	&\mv_+(\xv_{\ast}|\tilde{X}^{\mathrm{in}},\tilde{X}^{\mathrm{out}}) = (\mathbf{k}(\xv_\ast, X^{\mathrm{in}}) \otimes I_{n})K_{\mathrm{md}}^{-1}X^\mathrm{out} \label{eq:post:mean:vec}\\
	&\Sigma_+(\xv_{\ast}|\tilde{X}^{\mathrm{in}},\tilde{X}^{\mathrm{out}})=  	k(\xv_{\ast}, \xv_{\ast}) \otimes I_{n} \nonumber \\
	&\quad - (\mathbf{k}(\xv _\ast, X^{\mathrm{in}}) \otimes I_{n})K_{\mathrm{md}}^{-1} (\mathbf{k}(\xv_\ast, X^{\mathrm{in}}) \otimes I_{n})^\top.
\end{align}

The above derivation focuses once again on one single trajectory as offline data. If multiple trajectories have been collected, no covariance is needed at the transition between the different trajectories, as in the previous section.

\begin{remark}
	\label{rmk:correlations:components}
	In this paper, we assume independence among the different GPs modeling the different components of the unknown function $\mathbf{f}$. One interesting subject for future work is to omit this assumption. In this case, one could combine the here proposed approach with an intrinsic coregionalization method or a linear model of coregionalization \cite{alvarez2012kernels}.
\end{remark}

\subsection{Application to batch reactor, two-link planar robot, and cart-pole system}
We evaluate our approach in several numerical examples. For all numerical examples, we use a zero prior mean and a squared exponential kernel. For space reasons, we only explain the simulation setting in detail for the first example. We consider the following dynamics
\begin{subequations}
	\begin{align*}
		x_1(t+1) &= x_1(t) + T(-2c_1x_1^2(t)+2c_2x_2(t)) + w_1(t)\\
		x_2(t+1) &= x_2(t) + T(c_1x_1^2(t)- c_2x_2(t)) + w_2(t),
	\end{align*}
\end{subequations}
which corresponds to a discretized batch reactor \cite{Rawlings2017}. We consider $T = 0.1$, $c_1 = 0.16$, $c_2  = 0.0064$, normally distributed process noise with mean $\mu_w = 0$ and variance $\sigma^2_w = 10^{-6}I_n$, and normally distributed measurement noise with mean $\mu_r = 0$ and different variances as shown in Figure~\ref{fig:md}. We collect three trajectories containing 50 samples. 

Next, we test our approach for two four-dimensional systems with highly complex nonlinear dynamics. We consider a two-link planar robot with the dynamics and numerical parameter values as given in \cite{buisson2020actively} and a cart-pole system with the numerical parameters values from \cite{Barto1983}. The considered measurement noise variances are illustrated in Figure~\ref{fig:md} (middle and right plot). In both cases, we collect three trajectories containing 50 samples. 

In all examples, we use five iterations of slope/hyperparameter computations, see \cite{mchutchon2015nonlinear}. Furthermore, we implement a standard GP and the method proposed by \cite{mchutchon2015nonlinear} (by assuming that the different components are independent). We evaluate the performance for $N_\star=500$ random test data points sampled from a uniform distribution over some operating region of interest. More details can be found in the code of the simulations, which is provided under the link in footnote~3.

From Figure~\ref{fig:md}, one can see that the method proposed in this paper again outperforms the alternatives in terms of the MSE in all tested setting. Overall, the difference is more pronounced for larger noise levels. Furthermore, we can observe that the extension by \cite{mchutchon2015nonlinear} performs slightly better compared to the scalar case presented in the previous section. A reason for this observation may be that the extension proposed by \cite{mchutchon2015nonlinear} allows to learn the regression input noise variance using all outputs, which is not possible for a standard GP.

\section{CONCLUSION}
\label{sec:conclusion}
In this work, we analyzed the impact of regression input noise in case of dynamical systems modeled by Gaussian processes and introduced approaches to account for this noise in case of scalar and multidimensional systems. In several numerical examples, we showed that the consideration of the proposed extension substantially improves the performance compared to the state-of-the-art approaches. We expect that the method proposed in this paper will be beneficial for designing GP-based controllers and state estimators for nonlinear dynamical systems with improved performance.

Several topics are left for future research. One could refine the framework by using second order approximations (as also suggested by \cite{mchutchon2015nonlinear}), which is likely to improve the performance further, although inducing a larger computational complexity. Furthermore, one could investigate (theoretically and in simulations) the differences between the here proposed method and the (variational) methods suggested in \cite{doerr2018probabilistic,frigola2013bayesian}.

\bibliographystyle{IEEEtran}
\bibliography{IEEEabrv,input_noise_bib}

\end{document}